\documentclass[submission,copyright,creativecommons]{eptcs}
\providecommand{\event}{NCMA 2025} 

\usepackage{tikz}

\pdfoutput=1

\usepackage{iftex}

\ifpdf
  \usepackage{underscore}         
  \usepackage[T1]{fontenc}        
\else
  \usepackage{breakurl}           
\fi

\usepackage[english]{babel}

\usepackage{amssymb}
\usepackage{amsmath}
\usepackage{amsfonts}

\newcommand{\Accept}{{\sf Accept}}
\newcommand{\Reject}{{\sf Reject}}

\newcommand{\N}{\mathbb{N}}

\newcommand{\beginproof}{{\noindent \bf Proof.~}}
\newcommand{\myendproof}{\hspace*{\fill} $\Box$ \vspace{+0.2cm}}

\newtheorem{lemma}{Lemma}{\bf}{\it}
\newtheorem{proposition}[lemma]{Proposition}{\bf}{\it}
\newtheorem{theorem}[lemma]{Theorem}{\bf}{\it}
\newtheorem{corollary}[lemma]{Corollary}{\bf}{\it}
\newtheorem{definition}[lemma]{Definition}{\bf}{\it}
{\bf}{\rm}
{\bf}{\rm}
{\bf}{\it}

\title{On Repetitive Finite Automata\\ with Translucent Words}

\author{Franti{\v s}ek Mr\'az
\institute{%
Faculty of Mathematics and Physics\\
Charles University, Malostransk\'e n\'am.~25\\
118 00 Praha 1, Czech Republic}
\email{frantisek.mraz@mff.cuni.cz}
\and
Friedrich Otto
\institute{
Fachbereich Elektrotechnik/Informatik\\
Universit\"at Kassel\\
34109 Kassel, Germany}
\email{f.otto@uni-kassel.de}\\
}

\def\titlerunning{On Repetitive Finite Automata With Translucent Words}
\def\authorrunning{F.~Mr\'az and F.~Otto}

\begin{document}

\maketitle

\begin{abstract}
We introduce and study the repetitive variants of
the deterministic and the nondeterministic finite automaton
with translucent words (DFAwtw and NFAwtw).
On seeing the right sentinel,
a repetitive NFAwtw need not halt immediately, accepting or rejecting,
but it may change into another state and continue with its computation.
We establish that a repetitive DFAwtw already accepts a language
that is not even semi-linear,
which shows that the property of being repetitive increases the expressive capacity
of the DFAwtw and the NFAwtw considerably.
\vspace{+2mm}

\end{abstract}

\section{Introduction}\label{sec0}
The deterministic and the nondeterministic finite automaton \emph{with translucent letters} (or DFAwtl and NFAwtl)
was introduced by Nagy and Otto in~\cite{otto185} (see also~\cite{otto274})
as a reinterpretation of certain cooperating distributed systems of a very restricted type of deterministic restarting automata.
For each state~$q$ of an NFAwtl, there is a set $\tau(q)$ of \emph{translucent letters},
which is a subset of the input alphabet that contains those letters that the automaton cannot see when it is in state~$q$.
Accordingly, in each step, the NFAwtl just reads (and deletes) the first letter from the left that it can see, that is,
which is not translucent for the current state.
It has been shown that the NFAwtls accept a class of semi-linear languages
that properly contains all rational trace languages,
whereas its deterministic variant, the DFAwtl,
is properly less expressive.
In fact, the DFAwtl accepts a class of languages that is incomparable to the rational trace languages
with respect to inclusion~\cite{otto176,NaOtLATA2011,otto195,otto206}.
In addition, while the obvious upper bound for the time complexity of the membership problem for a DFAwtl is ${\rm DTIME}(n^2)$,
an improved upper bound of ${\rm DTIME}(n \cdot \log n)$ is derived in~\cite{NagKov14}.

In~\cite{otto268}, the authors present
a variant of the finite automaton with translucent letters
which, after reading and deleting a letter, does not return its head to the left end of its tape,
but that rather continues from the position of the letter just deleted.
When the end-of-tape marker is reached,
this automaton can decide whether to accept, reject, or continue with its computation,
which means that it changes its state and again reads the remaining tape contents from the beginning.
The latter property of the automaton is called `repetitiveness'.
This type of automaton, called a \emph{non-returning finite automaton with translucent letters} or an \emph{nrNFAwtl},
is strictly more expressive than the NFAwtl.
This result also holds for the deterministic case, although
the deterministic variant, the \mbox{\emph{nrDFAwtl},}
is still not sufficiently expressive to accept all rational trace languages.

In~\cite{MrOtRAIROsub}, the nrDFAwtl and the nrNFAwtl are compared to the jumping finite automaton,
the right one-way jumping finite automaton of~\cite{BiHo_IC284,CFY_IJFCS27}, and the
right-revolving finite automaton of~\cite{BBHK_IC207}, deriving the complete taxonomy of the resulting
classes of languages.

While an NFAwtl halts immediately when it sees its end-of-tape marker,
either accepting or rejecting,
a non-returning NFAwtl as described above is repetitive, that is, it may continue its computation in the corresponding situation.
In~\cite{otto272}, the authors study the influence that this property has on automata with translucent letters.
As it turns out, NFAwtls that are repetitive are equivalent to NFAwtls that are non-repetitive,
while the repetitive DFAwtls are strictly more expressive than the DFAwtls that are not repetitive.
On the other hand, nondeterministic and deterministic finite automata with translucent letters that
are non-returning and non-repetitive accept just the regular languages.
That is, they are equivalent to finite automata without translucent letters.
A recent survey on the various types of automata with translucent letters can be found
in~\cite{otto269}.

Finally, in~\cite{otto271}, the finite automaton with translucent letters is generalized by
extending the sets of translucent letters
to sets of translucent words, which yields the \emph{finite automaton with translucent words}
or \emph{NFAwtw}.
An NFAwtw reads (and deletes) the first letter from the left that is only preceded
by a prefix that is a product of words that are translucent for the current state.
This gives the automaton more control over the structure of the prefix ignored in a transition
than for an NFAwtl.
In order to guarantee that the resulting computation relation of an NFAwtw
can be computed efficiently,
the following two technical restrictions have been placed on the set $\tau(q)$ of translucent words associated with
a state~$q$ of an NFAwtw~$A$:
\begin{itemize}
\item the set of translucent words $\tau(q)$ is a finite prefix code, and
\item no word in the set $\tau(q)$ may begin with a letter $a$
that the NFAwtw $A$ can read in state~$q$, that is, for which $A$ has a possible
transition of the form $q'\in\delta(q,a)$.
\end{itemize}
Together these restrictions imply that the first letter from the left that
an NFAwtw can read in a state~$q$ can be determined
by simply scanning the current tape contents letter by letter from left to right.
It turned out that there are languages that are accepted by deterministic finite automata with
translucent words (that is, by \emph{DFAwtws}), but that are not even accepted by any
nondeterministic finite automata
with translucent letters.

The finite automaton with translucent words can be parameterized by placing two
restrictions on the size of the sets of translucent words admitted:
\begin{enumerate}
 \item An NFAwtw $A$ is $k$-cardinality-restricted for some integer $k\ge 1$,
 if each set of translucent words of $A$ contains at most $k$ elements.
 \item An NFAwtw $A$ is $\ell$-length-restricted for some integer $\ell\ge 1$,
 if no set of translucent words of $A$ contains a word of length larger than~$\ell$.
\end{enumerate}
Obviously, the $1$-length-restricted NFAwtw is just the NFAwtl,
and moreover, the notion of cardinality-restriction carries over to the NFAwtl.
These two parameters induce infinite strictly ascending two-dimensional hierarchies
of language classes for the NFAwtw and as well as for the DFAwtw~\cite{NaOtTOCS}.
In fact, the hierarchy based on cardinality-restriction alone and the hierarchy based on length-restriction alone
both carry over to the case of binary alphabets~\cite{MrOtNov24}.

Here, we define and study the repetitive variants of the NFAwtw and its deterministic variant, the DFAwtw.
On seeing the end-of-tape marker,
such an automaton may either halt, accepting or rejecting,
or it may change its state and reposition its head on the first letter of the current tape contents,
continuing with its computation.

The following important results are derived:
\begin{itemize}
 \item There exists a repetitive DFAwtw that accepts a language which is not semi-linear (Theorem~\ref{ThmNonSemiLin}).
 \item There exists a language that is accepted by a repetitive NFAwtw, but not by any repetitive DFAwtw
 (Theorem~\ref{ThmLveeNotInRDFAwtw}).
 \item For repetitive DFAwtws, emptiness is undecidable (Theorem~\ref{ThmUndec}).
 Moreover, finiteness, regularity, inclusion, equivalence, and boundedness are undecidable for this type of automaton, too.
\end{itemize}
However,
closure and non-closure properties for the various classes of repetitive NFAwtws and DFAwtws
have not yet been determined.

\section{Definitions and Known Results on Finite Automata with Translucent Words}\label{sec1}

First we restate the definition of the finite automaton with translucent words
as defined in~\cite{otto271}.
However, we slightly change the definition by removing the final states and by adjusting
the definition of the transition function accordingly.
Here we use $\mathcal{P}(S)$ to denote the powerset of a set $S$ and
$\mathcal{P}_{\rm fin}(S)$ to denote the set of all finite subsets of~$S$.

\begin{definition}\label{DefNFAwtw}
A \emph{finite automaton with translucent words}, or an \emph{NFAwtw}, is defined by a 6-tuple
$$A = (Q,\Sigma,\lhd,\tau,I,\delta),$$
where $Q$ is a finite set of states,
$\Sigma$ is a finite input alphabet, $\lhd\not\in\Sigma$ is a special letter that serves as an end-of-tape marker,
$I\subseteq Q$ is a set of initial states,
$\tau: Q\to \mathcal{P}_{\rm fin}(\Sigma^*)$ is a \emph{translucency mapping},
and
$$\delta:Q\times (\Sigma\cup\{\lhd\}) \to \mathcal{P}(Q)\cup\{\Accept,\Reject\}$$
is a transition function.
Here we require that, for each state $q \in Q$ and each letter $a \in \Sigma$,
$\delta(q, a) \subseteq Q$ and $\delta(q,\lhd) \in \{\Accept,\Reject\}$.
The latter means
that, on seeing the sentinel~$\lhd$, the NFAwtw $A$ halts immediately, either accepting or rejecting.

For each state $q\in Q$, let $\Sigma_q^{(A)} = \{\,a\in\Sigma \mid \delta(q,a) \not= \emptyset\,\}$,
that is, $\Sigma_q^{(A)}$ contains those letters that $A$ can read in state~$q$.
It is required that the set of translucent words $\tau(q)$
satisfies the following two restrictions:
\begin{itemize}
 \item If $\tau(q)\not=\emptyset$, then $\tau(q)$ is a finite prefix code.
 \item No word in $\tau(q)$ begins with a letter from the set~$\Sigma_q^{(A)}$.
 \end{itemize}
Actually, this means that the set $\tau(q)\cup\Sigma_q^{(A)}$ is a finite prefix code.
\vspace{+2mm}

The computation relation $\vdash_A^*$ that $A$ induces on its set of configurations
$Q\cdot\Sigma^*\cdot\lhd \,\cup\,\{\Accept,\Reject\}$ is the reflexive and transitive
closure of the following single-step computation relation, where $q\in Q$ and $w\in\Sigma^*$:
$$qw\cdot\lhd \vdash_A \left\{ \begin{array}{ll}
     q'uv\cdot\lhd, & \mbox{if }w=uav,\,u\in(\tau(q))^*,\,a\in\Sigma_q^{(A)},\,v\in\Sigma^*,
                                \mbox{ and }q'\in\delta(q,a),\\
      \Reject,       & \mbox{if }w=uav,\, u\in(\tau(q))^*,\,a\in\Sigma\smallsetminus \Sigma_q^{(A)},\,v\in\Sigma^*,\mbox{ and }
      av\not\in\tau(q)\cdot\Sigma^*,\\
     \Accept,      & \mbox{if }w\in(\tau(q))^* \mbox{ and }\delta(q,\lhd) = \Accept,\\
     \Reject,      & \mbox{if }w\in(\tau(q))^*\mbox{ and }\delta(q,\lhd) = \Reject.
     \end{array}\right.
$$
A word $w\in\Sigma^*$ is \emph{accepted by} $A$ if there exist an initial state $q_0\in I$
and a computation $q_0w\cdot \lhd \vdash_A^* \Accept$.
Now
$L(A)$ denotes the \emph{language accepted by}~$A$ and
$\mathcal{L}({\sf NFAwtw})$ denotes the class of all languages
that are accepted by~NFAwtws.

An NFAwtw $A=(Q,\Sigma,\lhd,\tau,I,\delta)$ is \emph{deterministic} (or a \emph{DFAwtw}) if $|I|=1$ and
$|\delta(q,a)| \le 1$ for each $q\in Q$ and $a\in\Sigma$.
For a DFAwtw, we simply replace the set $I$ by the single initial state
and write $\delta(q,a) = q'$ instead of $\delta(q,a) = \{q'\}$.
Then $\mathcal{L}({\sf DFAwtw})$ denotes the class of all languages
that are accepted by DFAwtws.
\end{definition}

As $\tau(q)$ is a prefix code for each state~$q$,
the factorization $w=uav$, where $u\in(\tau(q))^*$, $a\in\Sigma$, $v\in\Sigma^*$,
and $av\not\in\tau(q)\cdot\Sigma^*$, is uniquely determined.
This is not the case without the requirement that $\tau(q)$ is a prefix code (see~\cite{otto271}).
From the definition of the single step computation relation,
we obtain the following property.

\begin{lemma}[\cite{otto271}]\label{LemAppend}
Let $A = (Q,\Sigma,\lhd,\tau,I,\delta)$ be an NFAwtw and assume that $quav\cdot\lhd \vdash_A puv\cdot\lhd$,
where $q,p\in Q$, $u\in(\tau(q))^*$, $a\in\Sigma_q^{(A)}$, and $v\in\Sigma^*$.
Then $quavw\cdot\lhd \vdash_A puvw\cdot\lhd$ for each word $w\in\Sigma^*$.
\end{lemma}

Let $D_1\subseteq \{a,b\}^*$ be the semi-Dyck language on $\Sigma=\{a,b\}$,
that is, $D_1$ is the language that is generated by the context-free grammar
$$G = (\{S\},\Sigma,S,\{S\to \lambda, S\to SS, S\to aSb\}).$$
Furthermore, let $\Gamma = \{a,b,c\}$, $\varphi:\Sigma^*\to \Gamma^*$
be the morphism that is defined through $a\mapsto ab$ and $b\mapsto c$,
and $L_1 = \varphi(D_1)$.

\begin{lemma}[\cite{otto271}]\label{LemL1}
The language $L_1$ is accepted by a DFAwtw, but not by any NFAwtl.
\end{lemma}

Each NFAwtl can be extended to an equivalent NFAwtl that only accepts after having read and deleted
its input completely (see, e.g., \cite{otto195}).
For NFAwtws,
the corresponding technical result holds as well.

\begin{lemma}[\cite{otto271}]\label{LemReadAll}
From a given NFAwtw $A$, one can construct an NFAwtw $B$ such that $L(B)=L(A)$ and~$B$ only accepts
once it has read and deleted its input completely.
\end{lemma}

The NFAwtw $B$ constructed in the proof of this result is inherently nondeterministic,
even if the given NFAwtw $A$ happens to be deterministic.
Based on this technical result, the following result has been derived.

\begin{proposition}[\cite{otto271}]\label{PropRegSubLAng}
If $A$ is an NFAwtw, then there exists a regular sublanguage $R$ of the language $L(A)$
such that $R$ is letter-equivalent to $L(A)$.
In fact, an NFA $B$ for the sublanguage $R$ can effectively be constructed from $A$.
\end{proposition}

Here two languages on the same alphabet are called \emph{letter-equivalent}
if they have identical images under the corresponding Parikh mapping (see, e.g.~\cite{otto195}).
This result has the following immediate consequence.

\begin{corollary}[\cite{otto271}]\label{CorSemiLin}
The language accepted by an NFAwtw is semi-linear, that is,
its Parikh image is a semi-linear subset of $\N^n$, where $n$ is the cardinality of the underlying alphabet.
\end{corollary}

In addition, Proposition~\ref{PropRegSubLAng} implies the following negative result,
where $L_{\rm lin}$ denotes the deterministic linear language
$L_{\rm lin} = \{\,a^nb^n\mid n\ge 0\,\}$.

\begin{proposition}[\cite{otto271}]\label{PropLlin}
$L_{\rm lin}\not\in\mathcal{L}({\sf NFAwtw})$.
\end{proposition}

Observe that $L_{\rm lin} = L_{\rm eq2} \cap (a^*\cdot b^*)$,
where $L_{\rm eq2} = \{\,w\in\{a,b\}^*\mid |w|_a = |w|_b\,\}\in\mathcal{L}(\mbox{\sf DFAwtl})$.
Thus, Proposition~\ref{PropLlin} implies, in particular, that
the language classes $\mathcal{L}({\sf DFAwtw})$ and $\mathcal{L}({\sf NFAwtw})$
are not closed under intersection and under intersection with regular languages.

Finally, the DFAwtws have been separated from the NFAwtws.
Let
$$L_\vee = \{\, w \in \Sigma^* \mid \exists n \ge 0: |w|_a = n \mbox{ and } |w|_b \in \{n, 2n\}\, \},$$
where $\Sigma=\{a,b\}$.
The language $L_\vee$ is a rational trace language, and hence, it is accepted by an NFAwtl,
but it is not accepted by any DFAwtl~\cite{otto206}.
In fact, $L_\vee$ is not even accepted by any DFAwtw, either.

\begin{theorem}[\cite{NaOtTOCS}]\label{ThmLvee}
$L_\vee \not\in \mathcal{L}(\mbox{\sf DFAwtw})$.
\end{theorem}

Hence, we have the following proper inclusion.

\begin{corollary}[\cite{otto271}]\label{CorDFAwtwInNFAwtw}
$\mathcal{L}({\sf DFAwtw}) \subsetneq \mathcal{L}({\sf NFAwtw})$.
\end{corollary}

\section{Repetitive Finite Automata with Translucent Words}\label{sec2}

Here we present the announced extension of the finite automaton with translucent words.

\begin{definition}\label{DefnrNFAwtw}
Let $A=(Q,\Sigma,\lhd,\tau,I,\delta)$ be an NFAwtw.

\begin{enumerate}
\item[{\rm (a)}]
The NFAwtw $A$ is called
\emph{repetitive} if,
for each state $q\in Q$, $\delta(q,\lhd)$ is either a subset of $Q$ or
$\delta(q,\lhd)\in\{\Accept,\Reject\}$.
We use {\sf RNFAwtw} $(${\sf RDFAwtw}$)$ to denote the class of repetitive NFAwtws $($DFAwtws$)$.
To distinguish the model of Definition~\ref{DefNFAwtw} from the repetitive NFAwtw,
the former is called \emph{non-repetitive}.

\item[{\rm (b)}] The (R)NFAwtw $A$
is \emph{$k$-cardinality-restricted} (or a \emph{$k$-r$($R$)$NFAwtw})
for some integer $k\ge 1$, if
$|\tau(q)|\le k$ for each state $q\in Q$.
If $A$ is deterministic, then it is called a \emph{$k$-r$($R$)$DFAwtw}.

\item[{\rm (c)}] The (R)NFAwtw $A$ is
\emph{$\ell$-length-restricted} (or an \emph{$\ell$-lr-$($R$)$NFAwtw})
for some integer $\ell\ge 1$,
if $|u|\le \ell$ for all $u\in\tau(q)$ and all $q\in Q$.
If $A$ is deterministic, then it is called an \emph{$\ell$-lr-$($R$)$DFAwtw}.
\end{enumerate}

\end{definition}

We now study the repetitive NFAwtw
and its deterministic counterpart.
The following technical result can be derived for the RNFAwtw in the same way as for the NFAwtw.

\begin{lemma}\label{LemReadAllRNFA}
From a given RNFAwtw $A$, one can construct an RNFAwtw $B$ such that $L(B)=L(A)$ and~$B$ only accepts
once it has read and deleted its input completely.
\end{lemma}

On the other hand, Lemma~\ref{LemAppend} cannot be extended to the RNFAwtw,
if the automaton changes its state at the right sentinel.
For example, assume that $\delta(q,\lhd)= \{q'\}$,
$\delta(q,a)=\emptyset$, and $\tau(q)= \{aa\}$,
where $q,q'$ are states of an RDFAwtw $A$ with the input alphabet~$\{a\}$.
Then, $qaa\cdot\lhd \vdash_{A} q'aa\cdot\lhd$, while $qaaa \cdot\lhd \vdash_{A} \Reject,$
since $\tau(q) = \{aa\}$ and $\delta(q,a)$ is empty.

The deterministic linear language
$$L_{\rm lin} = \{\,a^nb^n \mid n\ge 1\,\}$$
is not accepted by any NFAwtw~\cite{otto271}.
Below we shall see that this language is accepted by some repetitive NFAwtw.
However, it is still open whether or not this language is accepted by any RDFAwtw.

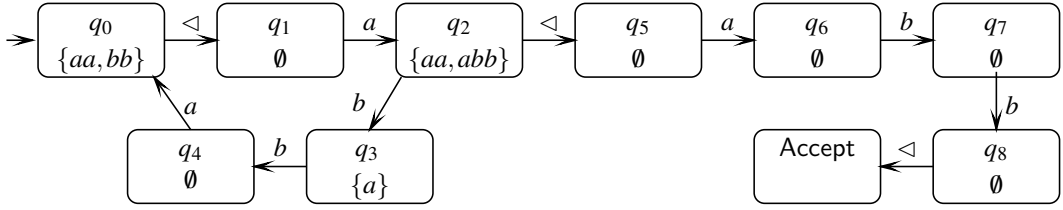
\begin{figure}
\centering
\ifpdf
\begin{tikzpicture}[x=0.70mm, y=0.70mm, inner xsep=0pt, inner ysep=0pt, outer xsep=0pt, outer ysep=0pt]
\path[line width=0mm] (43.00,50.00) rectangle +(204.00,42.00);
\definecolor{L}{rgb}{0,0,0}
\path[line width=0.21mm, draw=L] (51.00,78.00) .. controls (51.00,76.89) and (51.89,76.00) .. (53.00,76.00) .. controls (63.00,76.00) and (63.00,76.00) .. (73.00,76.00) .. controls (74.11,76.00) and (75.00,76.89) .. (75.00,78.00) .. controls (75.00,83.00) and (75.00,83.00) .. (75.00,88.00) .. controls (75.00,89.11) and (74.11,90.00) .. (73.00,90.00) .. controls (63.00,90.00) and (63.00,90.00) .. (53.00,90.00) .. controls (51.89,90.00) and (51.00,89.11) .. (51.00,88.00) .. controls (51.00,83.00) and (51.00,83.00) .. (51.00,78.00) -- cycle;
\draw(63.00,85.00) node[anchor=base]{\fontsize{9.96}{11.95}\selectfont $q_0$};
\draw(63.00,78.00) node[anchor=base]{\fontsize{9.96}{11.95}\selectfont $\{aa,bb\}$};
\path[line width=0.21mm, draw=L] (85.00,78.00) .. controls (85.00,76.89) and (85.89,76.00) .. (87.00,76.00) .. controls (97.00,76.00) and (97.00,76.00) .. (107.00,76.00) .. controls (108.11,76.00) and (109.00,76.89) .. (109.00,78.00) .. controls (109.00,83.00) and (109.00,83.00) .. (109.00,88.00) .. controls (109.00,89.11) and (108.11,90.00) .. (107.00,90.00) .. controls (97.00,90.00) and (97.00,90.00) .. (87.00,90.00) .. controls (85.89,90.00) and (85.00,89.11) .. (85.00,88.00) .. controls (85.00,83.00) and (85.00,83.00) .. (85.00,78.00) -- cycle;
\draw(97.00,85.00) node[anchor=base]{\fontsize{9.96}{11.95}\selectfont $q_1$};
\draw(97.00,78.00) node[anchor=base]{\fontsize{9.96}{11.95}\selectfont $\emptyset$};
\path[line width=0.21mm, draw=L] (119.00,78.00) .. controls (119.00,76.89) and (119.89,76.00) .. (121.00,76.00) .. controls (131.00,76.00) and (131.00,76.00) .. (141.00,76.00) .. controls (142.11,76.00) and (143.00,76.89) .. (143.00,78.00) .. controls (143.00,83.00) and (143.00,83.00) .. (143.00,88.00) .. controls (143.00,89.11) and (142.11,90.00) .. (141.00,90.00) .. controls (131.00,90.00) and (131.00,90.00) .. (121.00,90.00) .. controls (119.89,90.00) and (119.00,89.11) .. (119.00,88.00) .. controls (119.00,83.00) and (119.00,83.00) .. (119.00,78.00) -- cycle;
\draw(131.00,85.00) node[anchor=base]{\fontsize{9.96}{11.95}\selectfont $q_2$};
\draw(131.00,78.00) node[anchor=base]{\fontsize{9.96}{11.95}\selectfont $\{aa,abb\}$};
\path[line width=0.21mm, draw=L] (102.00,54.00) .. controls (102.00,52.89) and (102.89,52.00) .. (104.00,52.00) .. controls (114.00,52.00) and (114.00,52.00) .. (124.00,52.00) .. controls (125.11,52.00) and (126.00,52.89) .. (126.00,54.00) .. controls (126.00,59.00) and (126.00,59.00) .. (126.00,64.00) .. controls (126.00,65.11) and (125.11,66.00) .. (124.00,66.00) .. controls (114.00,66.00) and (114.00,66.00) .. (104.00,66.00) .. controls (102.89,66.00) and (102.00,65.11) .. (102.00,64.00) .. controls (102.00,59.00) and (102.00,59.00) .. (102.00,54.00) -- cycle;
\draw(113.50,61.00) node[anchor=base]{\fontsize{9.96}{11.95}\selectfont $q_3$};
\draw(114.00,54.00) node[anchor=base]{\fontsize{9.96}{11.95}\selectfont $\{a\}$};
\path[line width=0.21mm, draw=L] (68.00,54.00) .. controls (68.00,52.89) and (68.89,52.00) .. (70.00,52.00) .. controls (80.00,52.00) and (80.00,52.00) .. (90.00,52.00) .. controls (91.11,52.00) and (92.00,52.89) .. (92.00,54.00) .. controls (92.00,59.00) and (92.00,59.00) .. (92.00,64.00) .. controls (92.00,65.11) and (91.11,66.00) .. (90.00,66.00) .. controls (80.00,66.00) and (80.00,66.00) .. (70.00,66.00) .. controls (68.89,66.00) and (68.00,65.11) .. (68.00,64.00) .. controls (68.00,59.00) and (68.00,59.00) .. (68.00,54.00) -- cycle;
\draw(80.00,61.00) node[anchor=base]{\fontsize{9.96}{11.95}\selectfont $q_4$};
\draw(80.00,55.00) node[anchor=base]{\fontsize{9.96}{11.95}\selectfont $\emptyset$};
\path[line width=0.21mm, draw=L] (75.00,83.00) -- (85.00,83.00);
\definecolor{F}{rgb}{0,0,0}
\path[line width=0.21mm, draw=L, fill=F] (85.00,83.00) -- (82.20,83.70) -- (83.60,83.00) -- (82.20,82.30) -- (85.00,83.00) -- cycle;
\draw(80.00,85.00) node[anchor=base]{\fontsize{9.96}{11.95}\selectfont $\lhd$};
\path[line width=0.21mm, draw=L] (109.00,83.00) -- (119.00,83.00);
\path[line width=0.21mm, draw=L, fill=F] (119.00,83.00) -- (116.20,83.70) -- (117.60,83.00) -- (116.20,82.30) -- (119.00,83.00) -- cycle;
\draw(114.00,85.00) node[anchor=base]{\fontsize{9.96}{11.95}\selectfont $a$};
\path[line width=0.21mm, draw=L] (143.00,83.00) -- (153.00,83.00);
\path[line width=0.21mm, draw=L, fill=F] (153.00,83.00) -- (150.20,83.70) -- (151.60,83.00) -- (150.20,82.30) -- (153.00,83.00) -- cycle;
\draw(148.00,85.00) node[anchor=base]{\fontsize{9.96}{11.95}\selectfont $\lhd$};
\path[line width=0.21mm, draw=L] (102.00,59.00) -- (92.00,59.00);
\path[line width=0.21mm, draw=L, fill=F] (92.00,59.00) -- (94.80,58.30) -- (93.40,59.00) -- (94.80,59.70) -- (92.00,59.00) -- cycle;
\draw(97.00,61.00) node[anchor=base]{\fontsize{9.96}{11.95}\selectfont $b$};
\path[line width=0.21mm, draw=L] (120.00,76.00) -- (114.00,66.00);
\path[line width=0.21mm, draw=L, fill=F] (114.00,66.00) -- (116.04,68.04) -- (114.72,67.20) -- (114.84,68.76) -- (114.00,66.00) -- cycle;
\draw(112.00,69.50) node[anchor=base,rotate=0]{\fontsize{9.96}{11.95}\selectfont $b$};
\path[line width=0.21mm, draw=L] (80.00,66.00) -- (73.00,76.00);
\path[line width=0.21mm, draw=L, fill=F] (73.00,76.00) -- (74.03,73.30) -- (73.80,74.85) -- (75.18,74.11) -- (73.00,76.00) -- cycle;
\draw(80.00,69.00) node[anchor=base,rotate=0]{\fontsize{9.96}{11.95}\selectfont $a$};
\path[line width=0.21mm, draw=L] (153.00,78.00) .. controls (153.00,76.89) and (153.89,76.00) .. (155.00,76.00) .. controls (165.00,76.00) and (165.00,76.00) .. (175.00,76.00) .. controls (176.11,76.00) and (177.00,76.89) .. (177.00,78.00) .. controls (177.00,83.00) and (177.00,83.00) .. (177.00,88.00) .. controls (177.00,89.11) and (176.11,90.00) .. (175.00,90.00) .. controls (165.00,90.00) and (165.00,90.00) .. (155.00,90.00) .. controls (153.89,90.00) and (153.00,89.11) .. (153.00,88.00) .. controls (153.00,83.00) and (153.00,83.00) .. (153.00,78.00) -- cycle;
\draw(165.00,85.00) node[anchor=base]{\fontsize{9.96}{11.95}\selectfont $q_5$};
\draw(165.00,78.00) node[anchor=base]{\fontsize{9.96}{11.95}\selectfont $\emptyset$};
\path[line width=0.21mm, draw=L] (177.00,83.00) -- (187.00,83.00);
\path[line width=0.21mm, draw=L, fill=F] (187.00,83.00) -- (184.20,83.70) -- (185.60,83.00) -- (184.20,82.30) -- (187.00,83.00) -- cycle;
\draw(182.00,85.00) node[anchor=base]{\fontsize{9.96}{11.95}\selectfont $a$};
\path[line width=0.21mm, draw=L] (187.00,78.00) .. controls (187.00,76.89) and (187.89,76.00) .. (189.00,76.00) .. controls (199.00,76.00) and (199.00,76.00) .. (209.00,76.00) .. controls (210.11,76.00) and (211.00,76.89) .. (211.00,78.00) .. controls (211.00,83.00) and (211.00,83.00) .. (211.00,88.00) .. controls (211.00,89.11) and (210.11,90.00) .. (209.00,90.00) .. controls (199.00,90.00) and (199.00,90.00) .. (189.00,90.00) .. controls (187.89,90.00) and (187.00,89.11) .. (187.00,88.00) .. controls (187.00,83.00) and (187.00,83.00) .. (187.00,78.00) -- cycle;
\draw(199.00,85.00) node[anchor=base]{\fontsize{9.96}{11.95}\selectfont $q_6$};
\draw(199.00,78.00) node[anchor=base]{\fontsize{9.96}{11.95}\selectfont $\emptyset$};
\path[line width=0.21mm, draw=L] (211.00,83.00) -- (221.00,83.00);
\path[line width=0.21mm, draw=L, fill=F] (221.00,83.00) -- (218.20,83.70) -- (219.60,83.00) -- (218.20,82.30) -- (221.00,83.00) -- cycle;
\draw(216.00,85.00) node[anchor=base]{\fontsize{9.96}{11.95}\selectfont $b$};
\path[line width=0.21mm, draw=L] (221.00,78.00) .. controls (221.00,76.89) and (221.89,76.00) .. (223.00,76.00) .. controls (233.00,76.00) and (233.00,76.00) .. (243.00,76.00) .. controls (244.11,76.00) and (245.00,76.89) .. (245.00,78.00) .. controls (245.00,83.00) and (245.00,83.00) .. (245.00,88.00) .. controls (245.00,89.11) and (244.11,90.00) .. (243.00,90.00) .. controls (233.00,90.00) and (233.00,90.00) .. (223.00,90.00) .. controls (221.89,90.00) and (221.00,89.11) .. (221.00,88.00) .. controls (221.00,83.00) and (221.00,83.00) .. (221.00,78.00) -- cycle;
\draw(233.00,85.00) node[anchor=base]{\fontsize{9.96}{11.95}\selectfont $q_7$};
\draw(233.00,78.00) node[anchor=base]{\fontsize{9.96}{11.95}\selectfont $\emptyset$};
\path[line width=0.21mm, draw=L] (221.00,54.00) .. controls (221.00,52.89) and (221.89,52.00) .. (223.00,52.00) .. controls (233.00,52.00) and (233.00,52.00) .. (243.00,52.00) .. controls (244.11,52.00) and (245.00,52.89) .. (245.00,54.00) .. controls (245.00,59.00) and (245.00,59.00) .. (245.00,64.00) .. controls (245.00,65.11) and (244.11,66.00) .. (243.00,66.00) .. controls (233.00,66.00) and (233.00,66.00) .. (223.00,66.00) .. controls (221.89,66.00) and (221.00,65.11) .. (221.00,64.00) .. controls (221.00,59.00) and (221.00,59.00) .. (221.00,54.00) -- cycle;
\draw(233.00,61.00) node[anchor=base]{\fontsize{9.96}{11.95}\selectfont $q_8$};
\draw(233.00,54.00) node[anchor=base]{\fontsize{9.96}{11.95}\selectfont $\emptyset$};
\path[line width=0.21mm, draw=L] (233.00,77.00) -- (233.00,66.00);
\path[line width=0.21mm, draw=L, fill=F] (233.00,66.00) -- (233.70,68.80) -- (233.00,67.40) -- (232.30,68.80) -- (233.00,66.00) -- cycle;
\draw(236.00,69.00) node[anchor=base,rotate=0]{\fontsize{9.96}{11.95}\selectfont $b$};
\path[line width=0.21mm, draw=L] (187.00,54.00) .. controls (187.00,52.89) and (187.89,52.00) .. (189.00,52.00) .. controls (199.00,52.00) and (199.00,52.00) .. (209.00,52.00) .. controls (210.11,52.00) and (211.00,52.89) .. (211.00,54.00) .. controls (211.00,59.00) and (211.00,59.00) .. (211.00,64.00) .. controls (211.00,65.11) and (210.11,66.00) .. (209.00,66.00) .. controls (199.00,66.00) and (199.00,66.00) .. (189.00,66.00) .. controls (187.89,66.00) and (187.00,65.11) .. (187.00,64.00) .. controls (187.00,59.00) and (187.00,59.00) .. (187.00,54.00) -- cycle;
\draw(199.00,61.00) node[anchor=base]{\fontsize{9.96}{11.95}\selectfont $\Accept$};
\path[line width=0.21mm, draw=L] (221.00,59.00) -- (211.00,59.00);
\path[line width=0.21mm, draw=L, fill=F] (211.00,59.00) -- (213.80,58.30) -- (212.40,59.00) -- (213.80,59.70) -- (211.00,59.00) -- cycle;
\draw(216.00,61.00) node[anchor=base]{\fontsize{9.96}{11.95}\selectfont $\lhd$};
\path[line width=0.21mm, draw=L] (45.00,83.00) -- (50.00,83.00);
\path[line width=0.21mm, draw=L, fill=F] (50.00,83.00) -- (47.20,83.70) -- (48.60,83.00) -- (47.20,82.30) -- (50.00,83.00) -- cycle;
\end{tikzpicture}%
\else
\fi
\caption{\label{figa2nb2n}%
 The RDFAwtw $A_{\rm 2lin}$ for the language $L_{\rm 2lin} = \{\, a^{2n} b^{2n} \mid n \ge 1\,\}$.}
\end{figure}

\begin{lemma}\label{PropL2lin}
The language $L_{\rm 2lin} = \{\,a^{2n}b^{2n}\mid n\ge 1\,\}$ is accepted by a repetitive DFAwtw.
\end{lemma}

\beginproof
Let $A_{\rm 2lin} = (Q,\{a,b\},\lhd,\tau,q_0,\delta)$, where $Q=\{q_0,q_1,\ldots,q_8\}$,
be the RDFAwtw that is described in Figure \ref{figa2nb2n}.
Here, in each node,
the associated set of translucent words $\tau(q)$ is written under the name of the state $q\in Q$,
and there is an oriented edge from a state $q$ to a state $q'$
that is labeled with a letter $x \in \{a,b,\lhd\}$, if $\delta(q,x) = q'$.
Of course, $\delta$ is undefined for all other pairs from $Q\times\{a,b\}$.
It can be checked that $A_{\rm 2lin}$ accepts the language $L_{\rm 2lin}$.
\myendproof

\begin{figure}
\centering
\ifpdf
\begin{tikzpicture}[x=0.70mm, y=0.70mm, inner xsep=0pt, inner ysep=0pt, outer xsep=0pt, outer ysep=0pt]
\path[line width=0mm] (58.00,80.00) rectangle +(170.00,42.00);
\definecolor{L}{rgb}{0,0,0}
\path[line width=0.21mm, draw=L] (66.00,108.00) .. controls (66.00,106.89) and (66.89,106.00) .. (68.00,106.00) .. controls (78.00,106.00) and (78.00,106.00) .. (88.00,106.00) .. controls (89.11,106.00) and (90.00,106.89) .. (90.00,108.00) .. controls (90.00,113.00) and (90.00,113.00) .. (90.00,118.00) .. controls (90.00,119.11) and (89.11,120.00) .. (88.00,120.00) .. controls (78.00,120.00) and (78.00,120.00) .. (68.00,120.00) .. controls (66.89,120.00) and (66.00,119.11) .. (66.00,118.00) .. controls (66.00,113.00) and (66.00,113.00) .. (66.00,108.00) -- cycle;
\draw(78.00,115.00) node[anchor=base]{\fontsize{9.96}{11.95}\selectfont $q_0$};
\draw(78.00,108.00) node[anchor=base]{\fontsize{9.96}{11.95}\selectfont $\emptyset$};
\path[line width=0.21mm, draw=L] (100.00,108.00) .. controls (100.00,106.89) and (100.89,106.00) .. (102.00,106.00) .. controls (112.00,106.00) and (112.00,106.00) .. (122.00,106.00) .. controls (123.11,106.00) and (124.00,106.89) .. (124.00,108.00) .. controls (124.00,113.00) and (124.00,113.00) .. (124.00,118.00) .. controls (124.00,119.11) and (123.11,120.00) .. (122.00,120.00) .. controls (112.00,120.00) and (112.00,120.00) .. (102.00,120.00) .. controls (100.89,120.00) and (100.00,119.11) .. (100.00,118.00) .. controls (100.00,113.00) and (100.00,113.00) .. (100.00,108.00) -- cycle;
\draw(112.00,115.00) node[anchor=base]{\fontsize{9.96}{11.95}\selectfont $q_1$};
\draw(112.00,108.00) node[anchor=base]{\fontsize{9.96}{11.95}\selectfont $\{aa,b\}$};
\path[line width=0.21mm, draw=L] (90.00,113.00) -- (100.00,113.00);
\definecolor{F}{rgb}{0,0,0}
\path[line width=0.21mm, draw=L, fill=F] (100.00,113.00) -- (97.20,113.70) -- (98.60,113.00) -- (97.20,112.30) -- (100.00,113.00) -- cycle;
\draw(95.00,115.00) node[anchor=base]{\fontsize{9.96}{11.95}\selectfont $a$};
\draw(129.00,115.00) node[anchor=base]{\fontsize{9.96}{11.95}\selectfont $\lhd$};
\path[line width=0.21mm, draw=L] (60.00,113.00) -- (66.00,113.00);
\path[line width=0.21mm, draw=L, fill=F] (66.00,113.00) -- (63.20,113.70) -- (64.60,113.00) -- (63.20,112.30) -- (66.00,113.00) -- cycle;
\path[line width=0.21mm, draw=L] (124.00,113.00) -- (134.00,113.00);
\path[line width=0.21mm, draw=L, fill=F] (134.00,113.00) -- (131.20,113.70) -- (132.60,113.00) -- (131.20,112.30) -- (134.00,113.00) -- cycle;
\path[line width=0.21mm, draw=L] (134.00,108.00) .. controls (134.00,106.89) and (134.89,106.00) .. (136.00,106.00) .. controls (146.00,106.00) and (146.00,106.00) .. (156.00,106.00) .. controls (157.11,106.00) and (158.00,106.89) .. (158.00,108.00) .. controls (158.00,113.00) and (158.00,113.00) .. (158.00,118.00) .. controls (158.00,119.11) and (157.11,120.00) .. (156.00,120.00) .. controls (146.00,120.00) and (146.00,120.00) .. (136.00,120.00) .. controls (134.89,120.00) and (134.00,119.11) .. (134.00,118.00) .. controls (134.00,113.00) and (134.00,113.00) .. (134.00,108.00) -- cycle;
\draw(146.00,115.00) node[anchor=base]{\fontsize{9.96}{11.95}\selectfont $q_2$};
\draw(146.00,108.00) node[anchor=base]{\fontsize{9.96}{11.95}\selectfont $\emptyset$};
\path[line width=0.21mm, draw=L] (158.00,113.00) -- (168.00,113.00);
\path[line width=0.21mm, draw=L, fill=F] (168.00,113.00) -- (165.20,113.70) -- (166.60,113.00) -- (165.20,112.30) -- (168.00,113.00) -- cycle;
\path[line width=0.21mm, draw=L] (168.00,108.00) .. controls (168.00,106.89) and (168.89,106.00) .. (170.00,106.00) .. controls (180.00,106.00) and (180.00,106.00) .. (190.00,106.00) .. controls (191.11,106.00) and (192.00,106.89) .. (192.00,108.00) .. controls (192.00,113.00) and (192.00,113.00) .. (192.00,118.00) .. controls (192.00,119.11) and (191.11,120.00) .. (190.00,120.00) .. controls (180.00,120.00) and (180.00,120.00) .. (170.00,120.00) .. controls (168.89,120.00) and (168.00,119.11) .. (168.00,118.00) .. controls (168.00,113.00) and (168.00,113.00) .. (168.00,108.00) -- cycle;
\draw(180.00,115.00) node[anchor=base]{\fontsize{9.96}{11.95}\selectfont $q_6$};
\draw(180.00,108.00) node[anchor=base]{\fontsize{9.96}{11.95}\selectfont $\emptyset$};
\path[line width=0.21mm, draw=L] (192.00,113.00) -- (202.00,113.00);
\path[line width=0.21mm, draw=L, fill=F] (202.00,113.00) -- (199.20,113.70) -- (200.60,113.00) -- (199.20,112.30) -- (202.00,113.00) -- cycle;
\draw(197.00,115.00) node[anchor=base]{\fontsize{9.96}{11.95}\selectfont $\lhd$};
\path[line width=0.21mm, draw=L] (202.00,108.00) .. controls (202.00,106.89) and (202.89,106.00) .. (204.00,106.00) .. controls (214.00,106.00) and (214.00,106.00) .. (224.00,106.00) .. controls (225.11,106.00) and (226.00,106.89) .. (226.00,108.00) .. controls (226.00,113.00) and (226.00,113.00) .. (226.00,118.00) .. controls (226.00,119.11) and (225.11,120.00) .. (224.00,120.00) .. controls (214.00,120.00) and (214.00,120.00) .. (204.00,120.00) .. controls (202.89,120.00) and (202.00,119.11) .. (202.00,118.00) .. controls (202.00,113.00) and (202.00,113.00) .. (202.00,108.00) -- cycle;
\draw(214.00,115.00) node[anchor=base]{\fontsize{9.96}{11.95}\selectfont $\Accept$};
\draw(209.00,108.00) node[anchor=base]{\fontsize{9.96}{11.95}\selectfont };
\draw(163.00,115.00) node[anchor=base]{\fontsize{9.96}{11.95}\selectfont $b$};
\path[line width=0.21mm, draw=L] (100.00,84.00) .. controls (100.00,82.89) and (100.89,82.00) .. (102.00,82.00) .. controls (112.00,82.00) and (112.00,82.00) .. (122.00,82.00) .. controls (123.11,82.00) and (124.00,82.89) .. (124.00,84.00) .. controls (124.00,89.00) and (124.00,89.00) .. (124.00,94.00) .. controls (124.00,95.11) and (123.11,96.00) .. (122.00,96.00) .. controls (112.00,96.00) and (112.00,96.00) .. (102.00,96.00) .. controls (100.89,96.00) and (100.00,95.11) .. (100.00,94.00) .. controls (100.00,89.00) and (100.00,89.00) .. (100.00,84.00) -- cycle;
\draw(111.50,91.00) node[anchor=base]{\fontsize{9.96}{11.95}\selectfont $q_4$};
\draw(111.50,84.00) node[anchor=base]{\fontsize{9.96}{11.95}\selectfont $\{a,bb\}$};
\draw(149.00,99.00) node[anchor=base]{\fontsize{9.96}{11.95}\selectfont $a$};
\path[line width=0.21mm, draw=L] (134.00,84.00) .. controls (134.00,82.89) and (134.89,82.00) .. (136.00,82.00) .. controls (146.00,82.00) and (146.00,82.00) .. (156.00,82.00) .. controls (157.11,82.00) and (158.00,82.89) .. (158.00,84.00) .. controls (158.00,89.00) and (158.00,89.00) .. (158.00,94.00) .. controls (158.00,95.11) and (157.11,96.00) .. (156.00,96.00) .. controls (146.00,96.00) and (146.00,96.00) .. (136.00,96.00) .. controls (134.89,96.00) and (134.00,95.11) .. (134.00,94.00) .. controls (134.00,89.00) and (134.00,89.00) .. (134.00,84.00) -- cycle;
\draw(146.00,91.00) node[anchor=base]{\fontsize{9.96}{11.95}\selectfont $q_3$};
\draw(146.00,84.00) node[anchor=base]{\fontsize{9.96}{11.95}\selectfont $\{aa,ab\}$};
\path[line width=0.21mm, draw=L] (146.00,106.00) -- (146.00,96.00);
\path[line width=0.21mm, draw=L, fill=F] (146.00,96.00) -- (146.70,98.80) -- (146.00,97.40) -- (145.30,98.80) -- (146.00,96.00) -- cycle;
\path[line width=0.21mm, draw=L] (66.00,84.00) .. controls (66.00,82.89) and (66.89,82.00) .. (68.00,82.00) .. controls (78.00,82.00) and (78.00,82.00) .. (88.00,82.00) .. controls (89.11,82.00) and (90.00,82.89) .. (90.00,84.00) .. controls (90.00,89.00) and (90.00,89.00) .. (90.00,94.00) .. controls (90.00,95.11) and (89.11,96.00) .. (88.00,96.00) .. controls (78.00,96.00) and (78.00,96.00) .. (68.00,96.00) .. controls (66.89,96.00) and (66.00,95.11) .. (66.00,94.00) .. controls (66.00,89.00) and (66.00,89.00) .. (66.00,84.00) -- cycle;
\draw(77.50,91.00) node[anchor=base]{\fontsize{9.96}{11.95}\selectfont $q_5$};
\draw(77.50,84.00) node[anchor=base]{\fontsize{9.96}{11.95}\selectfont $\{aa,ab\}$};
\path[line width=0.21mm, draw=L] (134.00,89.00) -- (124.00,89.00);
\path[line width=0.21mm, draw=L, fill=F] (124.00,89.00) -- (126.80,88.30) -- (125.40,89.00) -- (126.80,89.70) -- (124.00,89.00) -- cycle;
\path[line width=0.21mm, draw=L] (78.00,96.00) -- (78.00,106.00);
\path[line width=0.21mm, draw=L, fill=F] (78.00,106.00) -- (77.30,103.20) -- (78.00,104.60) -- (78.70,103.20) -- (78.00,106.00) -- cycle;
\draw(81.00,99.00) node[anchor=base]{\fontsize{9.96}{11.95}\selectfont $b$};
\draw(129.00,91.00) node[anchor=base]{\fontsize{9.96}{11.95}\selectfont $b$};
\path[line width=0.21mm, draw=L] (100.00,89.00) -- (90.00,89.00);
\path[line width=0.21mm, draw=L, fill=F] (90.00,89.00) -- (92.80,88.30) -- (91.40,89.00) -- (92.80,89.70) -- (90.00,89.00) -- cycle;
\draw(95.00,91.00) node[anchor=base]{\fontsize{9.96}{11.95}\selectfont $\lhd$};
\end{tikzpicture}%
\else
\fi
\caption{\label{figa2n1b2n1}%
 The RDFAwtw $A_{\rm 2lin1}$ for the language $L_{\rm 2lin1} = \{\, a^{2n+1} b^{2n+1} \mid n \ge 0\,\}$. }
\end{figure}
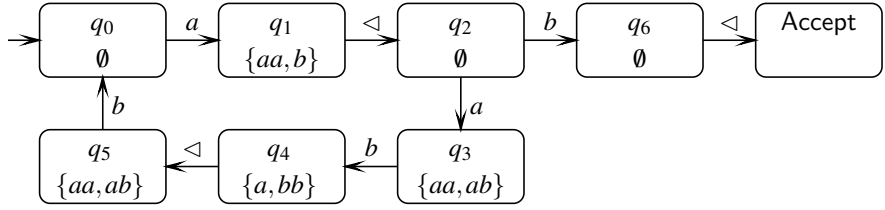

In essentially the same way, also the following result can be proved.
A corresponding automaton is presented in Figure~\ref{figa2n1b2n1}.

\begin{lemma}\label{PropL2n1lin}
The language $L_{\rm 2lin1} = \{\,a^{2n+1}b^{2n+1}\mid n\ge 0\,\}$ is accepted by a repetitive DFAwtw.
\end{lemma}

By forming the disjoint union of the RDFAwtws $A_{\rm 2lin}$ and $A_{\rm 2lin1}$, we obtain an RNFAwtw
for the language $L_{\rm lin}$, that is, we have the following consequence.

\begin{corollary}\label{CorLin}
The language $L_{\rm lin} = \{\,a^nb^n \mid n\ge 1\,\}$ is accepted by a repetitive NFAwtw.
\end{corollary}

Clearly, the language $L_{\rm 2lin}$ does not contain a regular sublanguage that is letter-equivalent
to the language itself, as, for each $n\ge 1$, $a^{2n}b^{2n}$ is the only word in $L_{\rm 2lin}$
that has length $4n$.
Hence,
by Proposition~\ref{PropRegSubLAng}, the language $L_{\rm 2lin}$ is not accepted by any NFAwtw.
In particular, this shows that Proposition~\ref{PropRegSubLAng} does not extend to the repetitive NFAwtw.

As stated in Corollary~\ref{CorSemiLin},
each language accepted by an NFAwtw
is necessarily semi-linear.
This is no longer true if we consider NFAwtws that are repetitive.

\begin{theorem}\label{ThmNonSemiLin}
There exists an RDFAwtw $A_{\rm ex}$ over a binary alphabet such that
the language $L(A_{\rm ex})$ is not semi-linear.
\end{theorem}

\beginproof
We define the RDFAwtw $A_{\rm ex}$ as $A_{\rm ex} = (Q,\Sigma,\lhd,\tau,q_0,\delta)$,
where
$$Q=\{q_0,q_1,q_2,q_3,q_4,q_5,q_6,q_7,q_f\},\; \Sigma = \{a,b\},$$
and the functions $\tau$ and $\delta$ are specified as follows:
$$\begin{array}{lcllcllcllcl}
\tau(q_0) & = & \{ab\}, & \tau(q_1) & = & \emptyset, & \tau(q_2) & = & \{bab\}, & \tau(q_3) & = & \emptyset,\\
\tau(q_4) & = & \{ab\}, & \tau(q_5) & = & \{ab\},    & \tau(q_6) & = & \emptyset,& \tau(q_7) & = & \emptyset,\\
\tau(q_f) & = & \emptyset,\\[+1mm]
\delta(q_0,\lhd) & = & q_1, &
\delta(q_1,a) & = & q_2, &
\delta(q_2,a) & = & q_2, &
\delta(q_2,\lhd) & = & q_3, \\
\delta(q_3,b) & = & q_4, &
\delta(q_4,b) & = & q_5, & \delta(q_4,\lhd) & = & q_6, &
\delta(q_5,b) & = & q_5, \\
\delta(q_5,\lhd) & = & q_1, & \delta(q_6,a) & = & q_7, & \delta(q_7,b) & = & q_f, & \delta(q_f,\lhd) & = & \Accept.
\end{array}$$

It can now be checked that $L(A_{\rm ex}) = \{\,(ab)^{2^n} \mid n\ge 1\,\} = L_{\rm ex}$.
For proving this result, we first establish the following technical statements.
\vspace{+2mm}

\noindent
{\bf Claim 1.} $abab = (ab)^{2^1} \in L(A_{\rm ex}).$
\vspace {+2mm}

\beginproof
Given the word $abab$ as input, the automaton $A_{\rm ex}$ executes the following computation:
$$\begin{array}{lclclclcl}
q_0abab\cdot\lhd & \vdash_{A_{\rm ex}} & q_1abab\cdot\lhd & \vdash_{A_{\rm ex}} & q_2bab\cdot\lhd & \vdash_{A_{\rm ex}} & q_3bab\cdot\lhd\\
                 & \vdash_{A_{\rm ex}} & q_4ab\cdot\lhd & \vdash_{A_{\rm ex}} & q_6ab\cdot\lhd & \vdash_{A_{\rm ex}} & q_7b\cdot\lhd\\
                 & \vdash_{A_{\rm ex}} & q_f\cdot\lhd & \vdash_{A_{\rm ex}} & \Accept.
\end{array}$$
\vspace{-7mm}

\myendproof

\vspace{+2mm}

\noindent
{\bf Claim 2.} For all $n\ge 2$, $q_1(abab)^n\cdot\lhd \vdash_{A_{ \rm ex}}^* q_3(bab)^n \vdash_{A_{\rm ex}}^* q_1(ab)^n\cdot\lhd.$
\vspace {+2mm}

\beginproof
We proceed by induction on $n$.
If $n=2$, then we obtain the following computation:
$$\begin{array}{lcllcllcllcl}
q_1(abab)^2\cdot\lhd & = & q_1abababab\cdot\lhd & \vdash_{A_{\rm ex}} & q_2bababab\cdot\lhd & \vdash_{A_{\rm ex}} & q_2babbab\cdot\lhd\\
                 & \vdash_{A_{\rm ex}} & q_3babbab\cdot\lhd & \vdash_{A_{\rm ex}} & q_4abbab\cdot\lhd
                 & \vdash_{A_{\rm ex}} & q_5abab\cdot\lhd \\
                 & \vdash_{A_{\rm ex}} & q_1abab\cdot\lhd & = & q_1(ab)^2\cdot\lhd.
\end{array}$$

For the general case, we consider the input $(abab)^{n+1} = abab(abab)^n$:
$$\begin{array}{lclclcl}
q_1abab(abab)^n\cdot\lhd & \vdash_{A_{\rm ex}} & q_2bab(abab)^n\cdot\lhd & \vdash_{A_{\rm ex}} & q_2babbab(abab)^{n-1}\cdot\lhd\\
                   & \vdash_{A_{\rm ex}}^{n-1} & q_2bab(bab)^{n}\cdot\lhd & \vdash_{A_{\rm ex}} & q_3(bab)^{n+1}\cdot\lhd \\
                   & \vdash_{A_{\rm ex}} & q_4ab(bab)^n\cdot\lhd & \vdash_{A_{\rm ex}} & q_5abab(bab)^{n-1}\cdot\lhd \\
                   & \vdash_{A_{\rm ex}}^{n-1} & q_5ab(ab)^n\cdot\lhd & \vdash_{A_{\rm ex}} & q_1(ab)^{n+1}\cdot\lhd.
\end{array}$$
\vspace{-7mm}

\myendproof

Together Claims 1 and 2 imply that $L_{\rm ex} \subseteq L(A_{\rm ex})$,
since, for each $n\ge 2$,
$$q_0(ab)^{2^n}\cdot \lhd \vdash_{A_{\rm ex}} q_1(abab)^{2^{n-1}}\cdot\lhd \vdash_{A_{\rm ex}}^* q_1abab\cdot\lhd \vdash_{A_{\rm ex}}^* \Accept.$$
\vspace{-5mm}

Conversely, assume that $w\in L(A_{\rm ex})$.
Then the computation of the automaton $A_{\rm ex}$ on the input~$w$ is accepting,
that is, it has the following form:
$$q_0w\cdot\lhd \vdash_{A_{\rm ex}} p_1w_1\cdot\lhd \vdash_{A_{\rm ex}} p_2w_2\cdot\lhd \vdash_{A_{\rm ex}} \cdots \vdash_{A_{\rm ex}}
p_tw_t\cdot\lhd  \vdash_{A_{\rm ex}} \Accept,$$
where $t\ge 1$, $p_1,p_2,\ldots,p_t\in Q$, and $w_1,w_2,\ldots,w_t\in\Sigma^*$.
From the definition of the functions $\tau$ and $\delta$, we see that
$p_1= q_1$ and that $w = (ab)^m$ for some $m\ge 0$,
$p_t= q_f$, and $w_t = \lambda$.
In fact, as
$$q_0\cdot\lhd  \vdash_{A_{\rm ex}} q_1\cdot\lhd \vdash_{A_{\rm ex}} \Reject$$
and
$$q_0ab\cdot\lhd \vdash_{A_{\rm ex}} q_1ab\cdot\lhd \vdash_{A_{\rm ex}} q_2b\cdot\lhd \vdash_{A_{\rm ex}} \Reject,$$
we can conclude that $m\ge 2$.
\vspace{+2mm}

\noindent
{\bf Claim 3.} For all $n\ge 1$, $(ab)^{2n+1} \not \in L(A_{\rm ex})$.
\vspace{+2mm}

\beginproof
For the input $(ab)^{2n+1}$, $A_{\rm ex}$ executes the  following computation:
$$\begin{array}{lclclclcl}
q_0ab(ab)^{2n}\cdot\lhd & \vdash_{A_{\rm ex}} &  q_1ab(ab)^{2n}\cdot\lhd & \vdash_{A_{\rm ex}} & q_2bab(ab)^{2n-2}ab\cdot\lhd\\
 &\vdash_{A_{\rm ex}}^{n-1} & q_2(bab)^{n}ab\cdot\lhd & \vdash_{A_{\rm ex}} & q_2(bab)^nb\cdot\lhd &
                \vdash_{A_{\rm ex}}  \Reject,
\end{array}$$
that is, $A_{\rm ex}$ rejects all uneven powers of $ab$.
\myendproof

Thus, it follows that $m$ is an even number.
Finally, assume that $m$ is not a power of two, that is, $m=2^k\cdot r$ for some $k\ge 1$ and an uneven number $r$.
Then, by Claims~2 and 3,
$$q_0(ab)^m\cdot\lhd  \vdash_{A_{\rm ex}} q_1(ab)^m\cdot\lhd = q_1(ab)^{2^k\cdot r}\cdot\lhd \vdash_{A_{\rm ex}}^*
q_1(ab)^r\cdot\lhd \vdash_{A_{\rm ex}}^* \Reject.$$
In summary, we have shown that $m=2^n$ for some integer $n\ge 1$,
that is, $w = (ab)^{2^n}$ is indeed an element of the language $L_{\rm ex}$.
It follows that $L(A_{\rm ex}) = L_{\rm ex}$, which completes the proof of Theorem~\ref{ThmNonSemiLin}.
\myendproof

As the language $L_{\rm ex}$ is not semi-linear, this gives the following result.

\begin{corollary}\label{CorNonSemiLin}
The language class $\mathcal{L}(\mbox{\sf RDFAwtw})$
contains languages that are not semi-linear.
\end{corollary}

As the NFAwtws only accept semi-linear languages, this also implies the following proper inclusions.

\begin{corollary}\label{CorInclRDFA}
$\mathcal{L}(\mbox{\sf DFAwtw}) \subsetneq \mathcal{L}(\mbox{\sf RDFAwtw})$
and
$\mathcal{L}(\mbox{\sf NFAwtw}) \subsetneq \mathcal{L}(\mbox{\sf RNFAwtw})$.
\end{corollary}

\section{Separating the RDFAwtw from the RNFAwtw}\label{SecSep}

The rational trace language
$$L_\vee = \{\, w \in \{a,b\}^* \mid \exists n \ge 0: |w|_a = n \mbox{ and } |w|_b \in \{n, 2n\}\, \}$$
is accepted by an NFAwtl, but according to Theorem~\ref{ThmLvee},
it is not accepted by any DFAwtw.
This means, in particular, that this language separates the DFAwtw from the NFAwtw (see Corollary~\ref{CorDFAwtwInNFAwtw}).
Here we prove that the language $L_\vee$ is not even accepted by any RDFAwtw.

\begin{theorem}\label{ThmLveeNotInRDFAwtw}
$L_\vee\not\in\mathcal{L}(\mbox{\sf RDFAwtw})$.
\end{theorem}

\beginproof
Assume to the contrary that there is an RDFAwtw
$A=(Q,\Sigma,\lhd,\tau,q_0,\delta)$ on $\Sigma=\{a,b\}$ such that $L(A) = L_\vee$,
and let
$$\ell = \max\{\,|u| \mid \exists q\in Q: u\in\tau(q)\,\} \mbox{ and } k = \max\{\,|\tau(q)| \mid q\in Q\,\},$$
that is, $A$ is $\ell$-length-restricted and $k$-cardinality-restricted.
Let $\Lambda>\ell$ be an integer that is sufficiently large.
In the following, we consider the accepting computations of $A$ for all inputs of the form
$a^nb^n$ and $a^nb^{2n}$, where $n\ge\Lambda$.

As $A$ is repetitive, it may have (one or more) states $q$ for which the set of letters $\Sigma_q^{(A)}$ is empty.
In fact, by introducing some additional states with this property, if necessary, we can assume, without loss of generality,
that, for each~$n\ge \Lambda$,
the accepting computation of $A$ on input $w_n = a^nb^n \in L_\vee$ has the following form:
$$\begin{array}{lcccccccccc}
q_0w_n\cdot\lhd & = & q_0a^nb^n\cdot\lhd & \vdash_A & p_0a^nb^n\cdot\lhd &
               \vdash_A & q_1z_1\cdot\lhd & \vdash_A & p_1z_1\cdot\lhd
                & \vdash_A & q_2z_2\cdot\lhd\\
                & \vdash_A & p_2z_2\cdot\lhd & \vdash_A & \ldots & \vdash_A & q_tz_t\cdot\lhd
               & \vdash_A & p_tz_t\cdot\lhd & \vdash_A & \Accept,
\end{array}$$
where, for all $i=0,1,2,\ldots,t$,
$q_i,p_i\in Q$, $\Sigma_{q_i}^{(A)} = \emptyset$, $\delta(q_i,\lhd) = p_i$,
$z_i\in\Sigma^{2n-i}$ is obtained from $w_n$ by reading and deleting $i$ letters,
$z_t\in(\tau(p_t))^*$, and $\delta(p_t,\lhd) = \Accept$.
In addition, $a^nb^n\in(\tau(q_0))^*$ and $z_i\in (\tau(q_i))^*$ for all $i=1,2,\ldots,t$.

As the set $\tau(q_0)$ is a finite prefix code and $a^nb^n\in(\tau(q_0))^*$,
we see that
$$\tau(q_0)\cap(a^*\cdot b^*) = \{a^{i_0},b^{j_0}\} \cup \{a^{r_1}b^{s_1},a^{r_2}b^{s_2},\ldots,a^{r_\nu}b^{s_\nu}\}$$
for some $1\le i_0,j_0\le \ell$, $\nu\ge 0$, $1\le r_1<r_2<\cdots<r_\nu< i_0$, and $s_1,s_2,\ldots,s_\nu\ge 1$
such that $r_\mu + s_\mu \le \ell$ for all $\mu=1,2,\ldots,\nu$.

Since $A$ is deterministic, $a^{m}b^m\in L_\vee$, and $a^mb^{2m}\in L_\vee$,
we can conclude that $a^mb^m\in(\tau(q_0))^*$ and $a^mb^{2m}\in (\tau(q_0))^*$ for each \mbox{$m\ge \Lambda$.}
Hence, for each $m \ge \Lambda$, there exist an index $f_m\in \{1,2,\ldots,\nu\}$ and integers $g_m,h_m\ge 0$
such that $m=g_m\cdot i_0 + r_{f_m} = h_m\cdot j_0 + s_{f_m}$.
As $1\le r_1<r_2<\cdots<r_\nu< i_0$,
it follows that $r_{f_m} \equiv m\!\!\!\mod i_0$ and
the index $f_m$ is uniquely determined by $i_0$ and $m$.
Analogously,
it follows that $2m = h_m'\cdot j_0 + s_{f_m}$ for some integer~$h_m'$,
which implies that $$m = 2m - m = h_m'\cdot j_0 +s_{f_m} - (h_m\cdot j_0 + s_{f_m}) = (h_m'-h_m) \cdot j_0.$$
Hence, each sufficiently large integer $m$ is necessarily a multiple of $j_0$, which means that $j_0 = 1$.
Moreover, as $m= g_m\cdot i_0 + r_{f_m}$, either $i_0=1$ and $\nu=0$, or
$i_0>1$, $\nu = i_0-1$, and $r_i = i$ for $i=1,2,\ldots,\nu$.
\vspace{+0mm}

If $\delta(p_0,a) = q_1$, then $z_1=a^{n-1}b^n$ is obtained from $w_n=a^nb^n$ by simply reading and deleting
the very first letter.
Accordingly, we obtain
$$q_0a^mb^m\cdot\lhd \vdash_A^2 q_1a^{m-1}b^m\cdot\lhd \mbox{ and } q_0a^mb^{2m}\cdot\lhd \vdash_A^2 q_1a^{m-1}b^{2m}\cdot\lhd$$
for all sufficiently large~$m$.
As $\Sigma_{q_1}^{(A)} = \emptyset$ and $\delta(q_1,\lhd) = p_1$,
we have
$$q_1a^{m-1}b^m\cdot\lhd \vdash_A p_1a^{m-1}b^m\cdot\lhd \mbox{ and } q_1a^{m-1}b^{2m}\cdot\lhd \vdash_A p_1a^{m-1}b^{2m}\cdot\lhd$$
for all sufficiently large~$m$.
Hence,
we can conclude, as above, that
$$\tau(q_1) \cap (a^*\cdot b^*) = \{a^{i_1},ab^{s'_1},a^2b^{s'_2},\ldots,a^{i_1-1}b^{s'_{i_1-1}},b\}$$
for some $i_1\ge 1$ and $s_1',s_2',\ldots,s'_{i_1-1}\ge 1$.
\vspace{+2mm}

If $\delta(p_0,a)$ is undefined and $\delta(p_0,b)=q_1$,
then $z_1=a^{n}b^{n-1}$ is obtained from $w_n=a^nb^n$ by reading and deleting an occurrence of the letter~$b$,
that is, a prefix of the form $a^nb^i$ of $a^nb^n$ is in the set $(\tau(p_0))^*$.
Again, as $a^mb^m,a^mb^{2m}\in L_\vee$, we see that
$$p_0a^mb^m\cdot\lhd \vdash_A q_1a^mb^{m-1}\cdot\lhd \mbox{ and }p_0a^mb^{2m}\cdot\lhd \vdash_A q_1a^mb^{2m-1}\cdot\lhd$$
for all sufficiently large~$m$.
Hence, we can conclude that
$$\tau(p_0) \cap (a^*\cdot b^*) = \{a^{i_1},ab^{s'_1},a^2b^{s'_2},\ldots,a^{i_1-1}b^{s'_{i_1-1}}\}$$
for some $i_1\ge 1$ and $s_1',s_2',\ldots,s'_{i_1-1}\ge 1$.
As $\Sigma_{q_1}^{(A)} = \emptyset$ and $\delta(q_1,\lhd) = p_1$,
we have
$$q_1a^{m}b^{m-1}\cdot\lhd \vdash_A p_1a^{m}b^{m-1}\cdot\lhd \mbox{ and } q_1a^{m}b^{2m-1}\cdot\lhd \vdash_A p_1a^{m}b^{2m-1}\cdot\lhd$$
for all sufficiently large~$m$,
which implies that
$$\tau(q_1) \cap (a^*\cdot b^*) = \{a^{i_2},ab^{s''_1},a^2b^{s''_2},\ldots,a^{i_2-1}b^{s''_{i_2-1}},b\}$$
for some $i_2\ge 1$ and $s_1'',s_2'',\ldots,s''_{i_2-1}\ge 1$.
\vspace{+2mm}

It follows that, for all sufficiently large values of~$m$, the accepting computations of $A$ on input $a^mb^m$
and on input $a^mb^{2m}$ consist of the exactly same sequence of transitional steps until the exponent
of one of the factors becomes small.
\vspace{+2mm}

Now consider a value of $n$ such that, for all $m\ge n$, the common initial part of all
the accepting computations of $A$ on input $a^mb^m$
and on input $a^mb^{2m}$ is of length $K>2\cdot |Q|$.
Then there are indices $0\le\alpha < \beta \le |Q|$ such that the states $p_\alpha$ and $p_\beta$ are identical.
Hence, for all $m\ge n$, we have the following accepting computations:
$$\begin{array}{llclclclcl}
q_0a^mb^m\cdot\lhd & \vdash_A^{2\cdot\alpha+1} & p_\alpha z_\alpha\cdot\lhd & \vdash_A^{2\cdot(\beta-\alpha)} & p_\beta z_\beta\cdot\lhd
                   & = & p_\alpha z_\beta\cdot\lhd &\vdash_A^* &\Accept
\end{array}$$
and
$$\begin{array}{llclclclcl}
q_0a^mb^{2m}\cdot\lhd & \vdash_A^{2\cdot\alpha+1} & p_\alpha z'_\alpha\cdot\lhd & \vdash_A^{2\cdot(\beta-\alpha)}
& p_\beta z'_\beta\cdot\lhd
                   & = & p_\alpha z'_\beta\cdot\lhd & \vdash_A^* & \Accept,
\end{array}$$
where $z_\alpha$ is obtained from $a^mb^m$ by reading and deleting $\alpha$ letters,
$z_\beta$ is obtained from $z_\alpha$ by reading and deleting further $\beta-\alpha$ letters,
$z'_\alpha$ is obtained from $a^mb^{2m}$ by reading and deleting $\alpha$ letters, and
$z'_\beta$ is obtained from $z'_\alpha$ by reading and deleting further $\beta-\alpha$ letters.
Thus,
$$z_\alpha = a^{m-i_\alpha}b^{m-j_\alpha},\, z_\beta = a^{m-i_\alpha-i_\beta}b^{m-j_\alpha-j_\beta},\,
z'_\alpha = a^{m-i_\alpha}b^{2m-j_\alpha},\, z'_\beta = a^{m-i_\alpha-i_\beta}b^{2m-j_\alpha-j_\beta}$$
for some integers $i_\alpha + j_\alpha = \alpha$ and $i_\beta+j_\beta = \beta-\alpha$.
\vspace{+2mm}

Consider now the input $a^{m+i_\beta}b^{m+j_\beta}$.
Then
$$\begin{array}{lclclclclcl}
q_0a^{m+i_\beta}b^{m+j_\beta}\cdot\lhd & \vdash_A^{2\cdot\alpha+1} & p_\alpha a^{m+i_\beta-i_\alpha}b^{m+j_\beta-j_\alpha}\cdot\lhd
& \vdash_A^{2\cdot(\beta-\alpha)} & p_\beta a^{m-i_\alpha}b^{m-j_\alpha}\cdot\lhd\\
                   & = & p_\alpha a^{m-i_\alpha}b^{m-j_\alpha}\cdot\lhd
                   & = & p_\alpha z_\alpha\cdot\lhd\\
                   & \vdash_A^* & \Accept.
\end{array}$$
As $m$ is large and $i_\beta,j_\beta \le \beta < |Q|$,
we see that $m+j_\beta < 2m < 2(m+i_\beta)$.
Hence, $a^{m+i_\beta}b^{m+j_\beta}\in L(A) = L_\vee$ implies
that $i_\beta = j_\beta$.
However, we also have the following computation:
$$\begin{array}{lclclclclcl}
q_0a^{m+i_\beta}b^{2m+j_\beta}\cdot\lhd & \vdash_A^{2\cdot\alpha+1} & p_\alpha a^{m+i_\beta-i_\alpha}b^{2m+j_\beta-j_\alpha}\cdot\lhd
& \vdash_A^{2\cdot(\beta-\alpha)} & p_\beta a^{m-i_\alpha}b^{2m-j_\alpha}\cdot\lhd\\
                   & = & p_\alpha a^{m-i_\alpha}b^{2m-j_\alpha}\cdot\lhd
                   & = & p_\alpha z'_\alpha\cdot\lhd\\
                   & \vdash_A^* & \Accept.
\end{array}$$
Now $m+i_\beta < 2m+j_\beta = 2m +i_\beta < 2m + 2i_\beta = 2\cdot(m+i_\beta)$ implies
that $a^{m+i_\beta}b^{2m+j_\beta}\not\in L_\vee$, a contradiction.
This proves that the language $L_\vee$ is not accepted by an RDFAwtw.
\myendproof

Hence, we have the following separation result.

\begin{corollary}\label{CorSepRDFAwtwFromRNFAwtw}
$\mathcal{L}(\mbox{\sf RDFAwtw}) \subsetneq \mathcal{L}(\mbox{\sf RNFAwtw}).$
\end{corollary}

\section{Emptiness Is Undecidable for RDFAwtws}\label{SecUndec}
From an NFAwtw $A$, an NFA $B$ can be constructed such that $L(B)$ is a sublanguage of $L(A)$
that is letter-equivalent to $L(A)$ (see Proposition~\ref{PropRegSubLAng}).
As the emptiness problem is decidable for NFAs (even in polynomial time),
and as $L(A)$ is empty if and only if $L(B)$ is empty, it thus follows that
the emptiness problem is decidable for NFAwtws.
In contrast to this fact, we now prove that this problem is undecidable for repetitive DFAwtws.
Our proof exploits a reduction from the \emph{Post Correspondence Problem} (PCP),
which can be stated as follows (see, e.g., \cite{HaKa97}):
\vspace{+0.2cm}

\noindent
\begin{tabular}{lcl}
{\bf Instance} & : & Two non-erasing morphisms $f,g:\Sigma^*\to\Delta^*$.\\
{\bf Question} & : & Is there a non-empty word $w\in\Sigma^+$ such that $f(w) = g(w)$?
\end{tabular}
\vspace{+0.2cm}

It is well-known that the PCP is undecidable in general, even when it is restricted to a binary alphabet~$\Delta$.

\begin{theorem}\label{ThmUndec}
The emptiness problem is undecidable for RDFAwtws.
\end{theorem}

\beginproof
Let $\Sigma=\{x_1,x_2,\ldots,x_m\}$ for some $m\ge 2$,
let $\Delta=\{a,b\}$, where
we can assume without loss of generality
that the two alphabets $\Sigma$ and $\Delta$ are disjoint,
and let $f,g:\Sigma^*\to\Delta^*$ be two non-erasing morphisms, that is,
$f(x_i) = u_i$ and $g(x_i) = v_i$ are non-empty words over~$\Delta$ for all $1\le i\le m$.

In addition, let $\Delta' = \{a',b'\}$ be a new alphabet such that $\Delta'$ is disjoint from $\Sigma$ and $\Delta$,
and let $\varphi':\Delta^* \to {\Delta'}^*$ be the morphism induced by mapping~$a$ to~$a'$ and~$b$ to~$b'$.
Finally, let $\Omega = \Sigma\cup\Delta\cup\Delta'$,
let $\pi_a: \Omega^* \to \Delta^*$ be the projection from $\Omega^*$ onto $\Delta^*$,
and let $\pi': \Omega^* \to \Delta^*$ be the morphism that is defined through
$\pi'(x_i) = \lambda$ for all $1\le i\le m$,
$\pi'(a) = \pi'(b) = \lambda$, and $\pi'(a') = a$ and $\pi'(b') = b$.
Then $\varphi' \circ \pi'$ is the projection from $\Omega^*$ onto ${\Delta'}^*$.

We now define an RDFAwtw $A_{(f,g)} = (Q,\Omega,\lhd,\tau,q_0,\delta)$ by taking
$$Q = \{q_0,q_1,q_2\} \cup \bigcup_{i=1}^m\left(\{\,p_y^{(i)} \mid y\mbox{ is a proper prefix of }u_i\,\} \cup
\{\,q_y^{(i)} \mid y\mbox{ is a proper prefix of }v_i\,\}\right)$$
and by defining the functions $\tau$ and $\delta$ as follows,
where ${\mathrm{pref}}(u_i,j)$ denotes the prefix of $u_i$ of length~$j$,
and ${\mathrm{pref}}(v_i,j)$ denotes the prefix of $v_i$ of length~$j$:
$$\begin{array}{@{\,}l@{\ \,}cllcllcl@{}}
\tau(q_0) & = & \Sigma \cup \{aa',bb'\},\\
\tau(q_1) & = & \emptyset, \\
\tau(q_2) & = & \emptyset,\\
\tau(p_y^{(i)}) & = & \multicolumn{7}{l}{\Sigma\cup \{a',b'\} \mbox{ for all }p_y^{(i)}\in Q,}\\
\tau(q_y^{(i)}) & = & \multicolumn{7}{l}{\Sigma\cup \{a,b\} \mbox{ for all }q_y^{(i)}\in Q,}\\[+2mm]
\delta(q_0,\lhd) & = & q_1,\\
\delta(q_1,x_i) & = & \multicolumn{7}{l}{p^{(i)}_{\lambda} \mbox{ for }1\le i\le m,}\\
\delta(p^{(i)}_{{\mathrm{pref}}(u_i,j)},a) & = & \multicolumn{7}{l}{p^{(i)}_{{\mathrm{pref}}(u_i,j+1)}
\mbox{ for }1\le i\le m \mbox{ and }0\le j < |u_i|-1, \mbox{ if }{\mathrm{pref}}(u_i,j+1) = {\mathrm{pref}}(u_i,j)a,}\\
\delta(p^{(i)}_{{\mathrm{pref}}(u_i,j)},b) & = & \multicolumn{7}{l}{p^{(i)}_{{\mathrm{pref}}(u_i,j+1)}
\mbox{ for }1\le i\le m \mbox{ and }0\le j < |u_i|-1, \mbox{ if }{\mathrm{pref}}(u_i,j+1) = {\mathrm{pref}}(u_i,j)b,}\\
\delta(p^{(i)}_{{\mathrm{pref}}(u_i,|u_i|-1)},a) & = & \multicolumn{7}{l}{q^{(i)}_{\lambda}
\mbox{ for }1\le i\le m, \mbox{ if }u_i = {\mathrm{pref}}(u_i,|u_i|-1)a,}\\
\delta(p^{(i)}_{{\mathrm{pref}}(u_i,|u_i|-1)},b) & = & \multicolumn{7}{l}{q^{(i)}_{\lambda}
\mbox{ for }1\le i\le m, \mbox{ if }u_i = {\mathrm{pref}}(u_i,|u_i|-1)b,}\\
\end{array}$$

$$\begin{array}{@{\,}l@{\ \,}cllcllcl@{}}
\delta(q^{(i)}_{{\mathrm{pref}}(v_i,j)},a') & = & \multicolumn{7}{l}{q^{(i)}_{{\mathrm{pref}}(v_i,j+1)}
\mbox{ for }1\le i\le m \mbox{ and }0\le j < |v_i|-1, \mbox{ if }{\mathrm{pref}}(v_i,j+1) = {\mathrm{pref}}(v_i,j)a,}\\
\delta(q^{(i)}_{{\mathrm{pref}}(v_i,j)},b') & = & \multicolumn{7}{l}{q^{(i)}_{{\mathrm{pref}}(v_i,j+1)}
\mbox{ for }1\le i\le m \mbox{ and }0\le j < |v_i|-1, \mbox{ if }{\mathrm{pref}}(v_i,j+1) = {\mathrm{pref}}(v_i,j)b,}\\
\delta(q^{(i)}_{{\mathrm{pref}}(v_i,|v_i|-1)},a') & = & \multicolumn{7}{l}{q_2
\mbox{ for }1\le i\le m, \mbox{ if }v_i = {\mathrm{pref}}(v_i,|v_i|-1)a,}\\
\delta(q^{(i)}_{{\mathrm{pref}}(v_i,|v_i|-1)},b') & = & \multicolumn{7}{l}{q_2
\mbox{ for }1\le i\le m, \mbox{ if }v_i = {\mathrm{pref}}(v_i,|v_i|-1)b,}\\
\delta(q_2,x_i) & = & \multicolumn{7}{l}{p^{(i)}_{\lambda} \mbox{ for all }1\le i\le m,}\\
\delta(q_2,\lhd) & = & \Accept,
\end{array}$$
and $\delta$ is undefined for all other pairs from $Q\times\Omega$.

It can now be verified that the language $L(A_{(f,g)})$ is non-empty if and only if the instance $(f,g)$ of the PCP
has a solution.
In fact, let $\psi_2:\Delta^* \to (\Delta\cup\Delta')^*$ be the morphism that is defined through
$a\mapsto aa'$ and $b\mapsto bb'$.
It can be checked that the language $L(A_{(f,g)})$ contains
some words from the shuffle of $x_{i_1}x_{i_2}\cdots x_{i_r}$ and $\psi_2(f(x_{i_1}x_{i_2}\cdots x_{i_r}))$
for each solution $x_{i_1}x_{i_2}\cdots x_{i_r}$ of $(f,g)$.
\myendproof

If $x=x_{i_1}x_{i_2}\cdots x_{i_r}$ is a solution of the PCP instance $(f,g)$,
also~$x^n$ is a solution of $(f,g)$ for each $n\ge 2$.
Hence, it follows that the language $L(A_{(f,g)})$ is either empty or infinite,
and it is infinite if and only if $(f,g)$ has a solution.
This has the following consequence.

\begin{corollary}\label{CorUndec2}
The finiteness problem is undecidable for RDFAwtws.
\end{corollary}

An RDFAwtw for the empty language is easily obtained.
Accordingly, the undecidability of the emptiness problem implies the following undecidability results.

\begin{corollary}\label{CorUndec3}
The inclusion problem and the equivalence problem are undecidable for RDFAwtws.
\end{corollary}

Let $(f,g)$ be an instance of the PCP, and let $A_{(f,g)}$ be the resulting
RDFAwtw as constructed in the proof of Theorem~\ref{ThmUndec}.
Assume that the language $L(A_{(f,g)})$ is regular.
Then also the language
$$L_{(f,g)} = L(A_{(f,g)})\cap (\Sigma^*\cdot \{aa',bb'\}^*)$$
is regular.
It can be checked that
$L_{(f,g)}$ consists of all words of the form
$w\psi_2(w)$,
where $w\in\Sigma^+$ is a solution for the instance $(f,g)$ of the PCP.

Assume that $(f,g)$ admits a solution~$w\in\Sigma^+$.
Then, for all $n\ge 2$, also $w^n$ is a solution for $(f,g)$, that is, $w^n(\psi_2(w))^n \in L_{(f,g)}$.
Let $k$ be the number of states of a minimal DFA for the language $L_{(f,g)}$.
Now pumping arguments show that, for all $n>k$, there exists an integer $\mu$, $1\le \mu < k$,
such that $w^{n+\mu}(\psi_2(w))^n$ is an element of the language $L_{(f,g)}$.
However, this contradicts the above observation about the form of the elements of this set,
as $w\not=\lambda$.
It follows that the set $L_{(f,g)}$, and therewith the language $L(A_{(f,g)})$, is not regular
whenever $(f,g)$ has a solution.
As the empty set is regular, this yields the following undecidability result.

\begin{corollary}\label{CorUndec4}
The regularity problem is undecidable for RDFAwtws.
\end{corollary}

Finally, a language $L\subseteq \Gamma^*$ is called \emph{bounded}
if there exist finitely many non-empty words $w_1,w_2,\ldots,w_k\in\Gamma^*$
such that $L$ is contained in the regular language $w_1^*\cdot w_2^*\cdots w_k^*$.
Now the boundedness problem is the problem of deciding whether a given language $L$
is bounded.
A recent survey on the status of this problem for various types of automata can be found in~\cite{IbarraDLT2024}.
While it is still open whether or not the boundedness problem is decidable for DFAwtws,
we have the following undecidability result.

\begin{corollary}\label{CorUndec5}
The boundedness problem is undecidable for RDFAwtws.
\end{corollary}

\beginproof
Let $(f,g)$ be an instance of the PCP and let $A_{(f,g)}$ be the RDFAwtw obtained from $(f,g)$
as in the proof of Theorem~\ref{ThmUndec}.
We now modify this RDFAwtw as follows.

Let $\Gamma = \{c,d\}$ be a new alphabet that is disjoint from $\Omega$,
let $\Omega' = \Omega \cup \Gamma$,
let $q_3$ be a new state, and let the functions~$\tau$ and $\delta$ be modified as follows:
$$\arraycolsep2pt
\begin{array}{lclclcl}
\tau'(q) & = & \left\{\begin{array}{ll}
                       \emptyset, & \mbox{if }q = q_3\\
                       \Sigma \cup \{aa',bb'\} \cup \Gamma, & \mbox{if }q=q_0,\\
                       \tau(q),   & \mbox{otherwise}
                      \end{array} \right\}
& \mbox{and} &
\delta'(q,x) & = & \left\{\begin{array}{ll}
                          q_3, &\mbox{if }q=q_2 \mbox{ and }x\in\Gamma\cup \{\lhd\},\\
                          q_3, &\mbox{if }q=q_3 \mbox{ and }x\in\Gamma,\\
                          \Accept, &\mbox{if }q=q_3 \mbox{ and }x=\lhd,\\
                          \delta(q,x), &\mbox{otherwise}.
                          \end{array}\right.
\end{array}$$
Let $A'_{(f,g)}$ be the new RDFAwtw.
If $(f,g)$ does not have a solution, then $L(A'_{(f,g)})$ is empty, and
hence, it is bounded.
However, if $(f,g)$ has a solution $w\in\Sigma^+$,
then $L(A'_{(f,g)})$ contains all words of the form $w\psi_2(w)z$, where $z\in\Gamma^*$,
which shows that this language is not bounded.
Thus, $L(A'_{(f,g)})$ is bounded if and only if $(f,g)$ does not have a solution.
As $A'_{(f,g)}$ is easily constructed from $(f,g)$, this yields the undecidability
of the boundedness problem.
\myendproof

\section{Conclusion}\label{SecCon}
We have shown that, by adding the property of repetitiveness,
the expressive capacity of the finite automata with translucent words is indeed severely extended.
However,
the following topics remain to be studied:

\begin{enumerate}
 \item The closure properties for the classes $\mathcal{L}(\mbox{\sf RDFAwtw})$ and $\mathcal{L}(\mbox{\sf RNFAwtw})$:
 It is easily seen that $\mathcal{L}({\sf RNFAwtw})$ is closed under union.
 On the other hand, the results on the language $L_\vee$ imply that the class
 $\mathcal{L}({\sf RDFAwtw})$ is neither closed under union nor under alphabetic morphisms.
 Moreover, by using the same proof idea as for NFAwtws, it can be shown that the complement
 of a language that is accepted by an RDFAwtw is accepted by an RNFAwtw.
 However, it remains open whether or not this deterministic class is closed under complementation.
 Finally, it is still open whether or not this class is closed under intersection (with regular languages).
 Obviously, it is closed under intersection with sets of the form $K^*$, where $K$ is a finite prefix code.

 \item What can we say about the complexity of the membership problem for an RDFAwtw?
 Obviously, this problem
 is decidable in quadratic time,
 but can we do better than that?

\end{enumerate}

\bibliographystyle{eptcs}
\bibliography{nrNFAwtwv2}

\end{document}